\documentstyle[12pt,epsfig]{article}

\large
\newcommand{\be}{\begin{eqnarray}}
\newcommand{\ee}{\end{eqnarray}}
\newcommand{\bea}{\left (\begin{array}{cc}}
\newcommand{\eea}{\right )\end{array}}

\newcommand{\mat}{\left ( \begin{array}{cc}}
\newcommand{\emat}{\end{array} \right )}
\newcommand{\matt}{\left ( \begin{array}{ccc}}
\newcommand{\ematt}{\end{array} \right )}
\newcommand{\matf}{\left ( \begin{array}{cccc}}
\newcommand{\ematf}{\end{array} \right )}
\newcommand{\vect}{\left ( \begin{array}{c}}
\newcommand{\evect}{\end{array} \right )}

\begin{document}
\setlength{\baselineskip}{17pt}
\pagestyle{empty}
\vfill
\eject
\begin{flushright}
SUNY-NTG-00/03
\end{flushright}

\vskip 2.0cm
\centerline{\Large \bf Chiral Random Matrix Model for Critical Statistics}

\vskip 1.2cm
\centerline{A.M. Garcia-Garcia and J.J.M. Verbaarschot}
\vskip 0.2cm
\centerline{\it
Department of Physics and Astronomy, SUNY, 
Stony Brook, New York 11794}
\vskip 1.5cm

\centerline{\bf Abstract}
We propose a random matrix model that interpolates between the chiral
random matrix ensembles and the chiral Poisson ensemble. By mapping this model 
on a non-interacting Fermi-gas we show that for energy differences less than 
a critical energy $E_c$ the spectral correlations are given by chiral Random 
Matrix Theory whereas for energy differences larger than $E_c$ the number 
variance shows a linear dependence on the energy difference 
with a slope that depends on the parameters of the model. 
If the parameters are scaled such that the slope remains fixed in the
thermodynamic limit, this model provides a description of QCD Dirac spectra in 
the universality class of  critical statistics. 
In this way a good description of QCD Dirac spectra for
gauge field configurations given by a liquid of instantons is obtained.


\vfill
\noindent

\eject
\pagestyle{plain}

\noindent
\section{ Introduction} 

By now it has been well-established that the smallest eigenvalues of the 
QCD Dirac operator are correlated according to a Random Matrix Theory
with the global symmetries of the QCD partition function
\cite{SV} (see \cite{tilojac,Kyoto} for recent reviews and a complete
list of references). In particular, this has been confirmed by the 
analysis of the low-energy effective theory \cite{Vplb,james,OTV,DOTV},
universality studies \cite{brezin,Damgaard,Sener1,GWu,Seneru}, lattice 
QCD simulations
\cite{HV,Vplb,berbenni,many,Damg99a,TiloSU3,Berb98c,Edwa99b,Berg99,Farc99a,Berb99}
and the study two-sublattice theories with disorder 
\cite{Gade,Simons,Taka,TTH}.
This means that the dynamical details of QCD are not important on 
energy scales of the order of the average level spacing. The natural question
that can be asked is at what  energy scale the dynamics of QCD becomes
relevant and how does this manifest itself in the Dirac spectrum.

The answer to this question has been understood  within the context
of effective theories \cite{GL,LS,Vplb,james,OTV,DOTV}. The effective
theory for the QCD Dirac spectrum is known as the partially quenched
effective partition function and was originally introduced to study the 
quenched approximation in QCD \cite{qChPT,pqChPT}. The central observation is
that in the domain where the mass dependence of the effective partition
function is given by the contribution of the constant fields (the zero
momentum modes) the Dirac eigenvalues are correlated according to chiral
Random Matrix Theory.
The relevant mass scale can thus be identified as the
scale for which the Compton wavelength of the lightest particle 
becomes equal to the size
of the box i.e., $M\sim 1/L_s$. In QCD, 
in the phase  of spontaneously broken chiral symmetry,
the lightest particles are the Goldstone modes with a mass
 given by $M^2 \sim K m$ (with $m$ the quark mass and  
$K = \Sigma/F^2$ in terms of the 
pion decay constant $F$ and the chiral condensate $\Sigma$).
The critical scale is thus given by \cite{Vplb}
\be
E_c = \frac 1{KL^2_s}.
\label{Thouless}
\ee
In the context of disordered condensed matter
systems \cite{Altshuler, HDgang,Montambaux} this energy scale 
is known as the Thouless
energy and also in this article we  will adopt this name.

A more intuitive interpretation of the Thouless energy 
has been given in the theory of mesoscopic systems \cite{Altshuler}.
The time scale $\hbar/E_c$ is the time for which an initially localized 
wave packet diffuses all over
space. For this reason 
the eigenvalues are correlated according to Random Matrix Theory  
for energy differences below $E_c$ (known as the ergodic regime).
At shorter time scales, different wave functions do not necessarily overlap
resulting in a weakening of correlations of the corresponding 
eigenvalues. For energy differences beyond the inverse elastic 
collision time $\tau_e$ the corresponding eigenvalues are completely uncorrelated 
(the Poisson ensemble). The domain inbetween $E_c$ and $\hbar/\tau_e$ is known as
the diffusive or Altshuler-Shklovskii domain. 
A third energy scale  is the average 
level spacing $\Delta \lambda$. The ratio $E_c/\Delta
\lambda$ is identified in mesoscopic physics as the dimensionless conductance.
It is equal to the number of subsequent levels correlated according to Random 
Matrix Theory. The existence of these domains has been confirmed
by numerical simulations of the Anderson model \cite{braun}.

In QCD the average level spacing is related to the order parameter
of the chiral phase transition, the chiral condensate, 
by the Banks-Casher formula \cite{BC} according to
$\Delta \lambda = \pi/\Sigma V$ (with $V$ the volume of Euclidean 
space-time). The prediction from (\ref{Thouless}) 
is that the
number of eigenvalues correlated according to chRMT is of the order
$F^2 \sqrt V$.

For increasing disorder the number of subsequent eigenvalues 
described by Random Matrix Theory decreases. For strong disorder we
expect that all states become localized with uncorrelated eigenvalues.
We thus expect a critical value of the disorder for which the three
scales, $\Delta \lambda$, $E_c$ and $ \hbar/\tau_e$ coincide. 
In particular, the dimensionless conductance becomes volume 
independent \cite{aronov}.
It has been conjectured \cite{shklov}
that at this point the eigenvalue correlations
are described by a new universality class known as 
critical statistics (see \cite{mirlin} for a review). In this class,
only the short range correlations 
of the eigenvalues 
are described by the usual random matrix ensembles
whereas the
number variance, $\Sigma^2(n)$, shows a linear $n$-dependence 
beyond this domain. What is relevant for QCD is that
such behavior has been observed in numerical
simulations of the 4-dimensional Anderson model \cite{zhar}.

The volume dependence of the Thouless energy has been investigated by means
of lattice QCD simulations \cite{Berb98c,TiloSU3,Berb99}
and instanton-liquid simulations \cite{james}. In essence, 
results from lattice QCD simulations are in complete agreement 
with theoretical results from partial quenched chiral perturbation theory. 
However, the results from instanton liquid simulations seem to deviate
from the prediction (\ref{Thouless}) with a scale independent constant
$K$; in that case
the Thouless energy only shows a weak
volume dependence. 
This raises the question whether the Dirac eigenvalues might be described
by critical statistics.
To address this issue we generalize a random matrix model for critical
statistics \cite{Moshe,kra1} to include the chiral symmetry of the
QCD partition function (section 2). In section 3 we map our model
on a partition function of noninteracting fermions. This model
is solved in the semi-classical limit in section 4, where we
obtain analytical expressions for the microscopic spectral density
and the two-point correlation function.
Comparisons
with instanton simulations are shown in section 5 and concluding
remarks are made in section 6.

\section{Definition of the Model}
The random matrix model of Moshe and Neuberger and Shapiro \cite{Moshe} 
is defined by the partition function
\be
Z =\int dH e^{-{\rm Tr} HH^\dagger} \int dU 
e^{-b {\rm Tr}( [U,H] [U,H]^\dagger)},
\label{MNS}
\ee
where $H$ is a  Hermitian and $U$  a Unitary $N\times N$ matrix. The
integration measures $dH$ and $dU$ are given by  the Haar measure. 
This model can be interpreted as the zero-dimensional limit of the 
Kazakov-Migdal model \cite{kazakov}.
It interpolates between the Gaussian Unitary Ensemble $(b=0$) 
and the Poisson Ensemble $(b \to \infty)$. Using the invariance
of the measure, the integral over $U$ can be replaced by an integral
over the eigenvalues of $U$. For $b \rightarrow \infty$ this partition
function is dominated by matrices
 $H$ that commute with  arbitrary 
diagonal unitary matrices. This set of matrices is 
the ensemble of diagonal Hermitian
matrices which is known as the Poisson ensemble. What is nice about this
model is that it preserves the unitary invariance which enables us
to take full advantage of the existing  random matrix theory methods.
In order to obtain a nontrivial $b$-dependence in the thermodynamic 
limit, the parameter $b$ has to be scaled as 
\be
b = h^2 N^2.
\ee
It has been shown that in this limit the model (\ref{MNS}) is 
equivalent \cite{kra1} to both a banded random  matrix
model \cite{band} with a power-like cutoff \cite{dittes}
and  to random matrix models with a
${\rm Tr}\log^2 H$ probability potential \cite{log2}. The correlation
functions of the latter model have been derived  by means of $q$-orthogonal
polynomials \cite{log2,chen} and Painlev\'e equations \cite{nishilog}.
  
In this paper we are interested in  chiral random matrix ensembles defined 
as ensembles 
of $N\times N $ random matrices with the structure
\be
D= \mat 0 & C \\ C^\dagger & 0 \emat,
\ee
where $C$ is an arbitrary complex matrix
$n\times (n+\nu) $ matrix ($N=2n+\nu$). 
Since the
matrix $D$ has exactly $\nu$ zero eigenvalues $ \nu$
is interpreted as the topological quantum number. The
 chiral Gaussian Unitary Ensemble (chGUE) with $N_f$ flavors 
is defined as the ensemble
of matrices $D$ with matrix elements $C$ distributed according to 
the Gaussian probability distribution
\be
P(C) \sim {\det}^{N_f} (D+m) e^{-N{\rm Tr} CC^\dagger}.
\label{pc}
\ee
Here, for simplicity we have taken all quark masses equal
to $m$. 
The probability distribution $P(C)$ has the unitary invariance
\be
C \rightarrow U C  V^{-1},
\label{decom}
\ee
where $U$ and $V$ are unitary matrices.
Since an arbitrary complex matrix can always be brought to diagonal form
by this transformation this invariance allows us to factorize
the probability distribution in a product over the eigenvalues of $C$
and the unitary matrices that diagonalize $C$. 

The generalization of the
model of Moshe, Neuberger and Shapiro to the chiral ensembles is immediate.
The interpolating model is defined by the partition function 
\be
Z_\nu = \int dC  {\det}^{N_f} (D+ m) 
e^{-\frac {\Sigma^2}{4h}{\rm Tr} D^\dagger D} \int d{\cal U} e^{-
\frac {\Sigma^2 h N^2}{4}
{\rm Tr} [D,{\cal U}] [D,{\cal U}]^\dagger},
\label{zdef}
\ee
where ${\cal U}$ has the chiral block structure
\be
{\cal U} = \mat U & 0 \\ 0 &V \emat,
\ee
with $U$ an $n\times n$ unitary matrix and $V$ an $(n+\nu)\times
(n+\nu)$ unitary matrix. If $\Sigma$ and $h$ are constants as in
(\ref{zdef}) we will refer to this model as the critical chiral
unitary ensemble. As we will see below, in order to make contact
with the Thouless energy in the partially quenched effective
partition function we have to scale $h$ with an additional factor
$1/\sqrt N$.  
The unitary invariance of this partition function
follows from the invariance of the Haar measure.
In comparison to \cite{Moshe}, an additional factor $\sim\Sigma^2/h$ has
been included in the probability distribution of the matrix elements.
As we will see below, this will guarantee that in the thermodynamic limit
the spectral density is $h$-independent to leading order in $h$. 
We will also find that
the partition function is normalized such that the $\Sigma$ represents
the chiral condensate by means of the Banks-Casher relation
$\Sigma = \pi \rho(0)/N$ (with $\rho(0)$ the spectral density around 
$\lambda = 0$).  

Decomposing into the blocks of $D$ and ${\cal U}$ the partition function can
be written as
\be
Z_\nu =m^{\nu}\int dC {\det}^{N_f} [C^\dagger C+m^2]
e^{-(1+2h^2N^2)\frac{\Sigma^2}{2h} {\rm Tr} CC^\dagger} \int dU dV
e^{\Sigma^2 h N^2
{\rm Re}{\rm Tr} U C V^{-1}C^\dagger}.
\label{ZUCVC}
\ee
The arbitrary complex matrix $C$ can be decomposed according to
\be
C = U_1 \Lambda U_2,
\ee
and the integral over $C$ can  be expressed as an
integral
over the eigenvalues $\Lambda_k$ and the unitary matrices  $U_1$ and
$U_2$. Up to a irrelevant constant, the Jacobian of this transformation 
is given by 
\be
J(\Lambda) =\Delta^2(\{\lambda_i^2\})
 \prod_k \lambda_k^{2\nu+1}.
\ee
where 
\be
\Delta(\{\lambda_i^2\}) =
\prod_{k<l}(\lambda_k^2 -\lambda_l^2)
\ee
is the Vandermonde determinant. Because of the unitary invariance,
the $U_1$ and $U_2$ dependence in the second exponent can be absorbed
in a redefinition of $U$and $V$, and  
the integrations over $U_1$ and $U_2$ just result
in an overall constant. Remarkably, the integral over $U$ and $V$ 
in (\ref{ZUCVC}) is an Itzykson-Zuber
type integral which is known analytically \cite{Russian,GW,Seneruv}
\be
\int dU dV e^{z{\rm Re}{\rm Tr} U \Lambda V^{-1} \Lambda } = 
 C \frac{ \det_{k,l} \left| I_\nu(z\lambda_k \lambda_l)\right|}
{ \Delta^2(\{\lambda_i^2\})\prod_{k=1}^{n} \lambda_k^{2\nu}}.
\label{IZ2}
\ee
The Vandermonde determinants cancel resulting in the partition function 
\be
Z_\nu = m^\nu \int  
\prod_{k=1}^{n} d\lambda_k \prod_{k=1}^n(\lambda_k^2 +m^2)^{N_f}
\det_{k,l} \left| (\lambda_k\lambda_l)^{\frac 12}
e^{-( \frac 12 +h^2N^2)\frac {\Sigma^2}{2h} (\lambda_k^2 +\lambda_l^2)}
I_\nu(h\Sigma^2N^2 \lambda_k \lambda_l)\right |  .\nonumber\\
\label{partigen}
\ee
The instanton-liquid Dirac spectra that  will be described by this model
were obtained in the quenched approximation and for zero
total topological charge. Therefore we will not 
attempt to solve this model for arbitrary $N_f$ but instead focus on
the technically simpler case of $N_f= 0$. Since the topological charge
does not give rise to additional complications we will consider the
case of arbitrary $\nu$. Below we will thus analyze the joint 
probability distribution
\be  
\rho_\nu(\lambda_1, \cdots, \lambda_n) \equiv  
\det_{k,l} \left| (\lambda_k\lambda_l)^{\frac 12}
e^{-( \frac 12 +h^2N^2)\frac {\Sigma^2}{2h} (\lambda_k^2 +\lambda_l^2)}
I_\nu(h\Sigma^2N^2 \lambda_k \lambda_l)\right |  .\nonumber\\
\label{parti}
\ee

\section{Fermi-Gas Representation of the Interpolating chiral Random 
Matrix Model}
In this section we rewrite  the joint probability distribution (\ref{parti}) 
in terms of the fermionic $n$-particle matrix element
\be
\rho_\nu(x_1, \cdots, x_n) \equiv  
 C\langle x_1 \cdots x_n| e^{-\beta H} |x_1 \cdots x_n \rangle,
\label{matrixelement}
\ee
where $H$ is the separable Hamiltonian
\be
H = \sum_k (-\partial_k^2 +\frac {4\nu^2-1}{4 x^2_k} + \omega^2 x_k^2),
\ee
and $C$ is an irrelevant constant.
The eigenfunctions of the single particle Hamiltonian are known in terms
of Laguerre polynomials. Specifically,
\be
(-\partial_x^2 +\frac {4\nu^2-1}{4 x^2} + \omega^2 x^2)\phi_n
= \omega_n \phi_n,
\ee
has the solutions
\be
\omega_n &=& (4n+2\nu + 2) \omega,\\
\phi_n &=&\frac{\sqrt 2 \omega^{1/4}}{ \sqrt h_n} e^{-\omega x^2/2} 
\left ( x\sqrt \omega \right )^{\nu +\frac 
12} L_n^\nu\left ( {x^2}\omega \right ).
\label{eigenfunctions}
\ee
The normalization factor $h_n = (n+\nu)!/n!$ ensures that the
 $\phi_n$ are normalized to unity
\be
\int_0^\infty \phi_n(x)\phi_n(x) dx = 1. 
\ee
Using completeness, the many-particle matrix element can be written as
\be
\langle x_1\cdots x_n| e^{-\beta H} |x_1\cdots x_n\rangle
= \det_{ij} \sum_n \phi_n(x_i) 
e^{-\beta\omega_n} \phi_n(x_j)
\label{me}
\ee
The sum over $n$ can be performed analytically using the identity
\be
\sum_{n=0}^\infty \frac{L_n^\nu(x) L_n^\nu(y)z^n}{h_n}=
\frac{(xyz )^{-\nu/2}}{1-z} \exp\left (-z\frac{x+y}{1-z}\right ) I_\nu 
\left (\frac{2\sqrt{xyz }}{1-z}\right).
\ee
With $z =\exp(-4\beta \omega)$ this results in
\be
\langle x_1\cdots x_n| e^{-\beta H} |x_1\cdots x_n\rangle
= \det_{ij} \left |\frac {\omega \sqrt{x_ix_j}}{\sinh 2\beta \omega} 
\exp\left (-\frac{\omega \cosh 2\beta \omega}{2\sinh 2\beta \omega}
(x^2_i+x^2_j)\right ) I_\nu 
\left (\frac {\omega}{\sinh 2\beta \omega} x_ix_j\right) \right |.
\nonumber \\
\ee 
Comparing this expression with (\ref{parti}) we find that
\be
\omega &=&   \sqrt{4h^2N^2 + 1}\frac {\Sigma^2}{2h},\\
\cosh 2\beta \omega &=& 1 + \frac 1{2h^2N^2}.
\ee

The joint probability distribution of
the eigenvalues is thus given by an $n-$particle diagonal matrix
element of the density operator. 
The average spectral density is equal to the average particle
density. It is obtained by integrating over the positions of all 
particles except one. The integral can be performed by 
rewriting the matrix elements (\ref{me}) 
in terms of a sum over permutations $\pi$ and $\pi'$,
\be
\frac 1{n!}\sum_{\omega_1< \cdots <\omega_n}\sum_{\pi\pi'}\sigma(\pi)
\sigma(\pi')
\phi_{\omega_{\pi(1)}}(x_1)\cdots \phi_{\omega_{\pi(n)}}(x_n)
e^{-\beta \sum_k \omega_k}
\phi_{\omega_{\pi'(1)}}(x_1)\cdots\phi_{\omega_{\pi'(n)}}(x_n),
\ee
where $\sigma(\cdot)$ is the sign of the permutation. Performing the
integrations over $x_2, \cdots,x_n$, by orthogonality we find that the
only nonzero contribution is for $\pi = \pi'$. Then the remaining sum
over $\pi(\omega_1)$ is just a sum over $\omega_1, \cdots, \omega_n$.
For the canonical ensemble
we thus find the one-particle density, 
\be
\langle \rho(x) \rangle\equiv \langle \sum_k \delta(x-x_k) \rangle =
 \frac 1{Z_n}\sum_{\omega_1< \cdots <\omega_N}
\sum_i|\phi_{\omega_i}(x_1)|^2e^{-\beta \sum_k \omega_k}.
\ee
In an occupation number representation this can be rewritten as
\be
\langle \rho(x) \rangle= \frac 1{Z_n}
\sum_{\sum_i n_i =n} \sum_i n_i |\phi_{\omega_i}(x)|^2
e^{-\beta \sum_k n_i\omega_k},
\ee
where the sum is over all $n_i\in\{0,\, 1\}$ subject to the condition given
below the summation sign and the canonical partition function is given by
\be
Z_n = \sum_{\sum_i n_i =n} 
e^{-\beta \sum_i n_i\omega_i}.
\ee

For completeness we give the following exact expressions for 
the  canonical partition function \cite{caselle,boulatov}  
\be
Z_n = \frac {e^{-2\beta \omega n^2 - 2\beta\nu \omega n}}
{(1-e^{-4\beta \omega})(1-e^{-8\beta\omega}) \cdots(1-e^{-4n\beta\omega})},
\ee
and the one-particle density 
\be
\langle \rho(x) \rangle= 
\sum_k |\phi_k(x)|^2\sum_{r=0}^\infty (-1)^r e^{-4\beta\omega(kr
-nr +\frac 12 r(r+1))}\prod_{p=N-r}^N(1-e^{-4\beta \omega p}).
\ee
They have been obtained by
writing the constraint $\sum_i n_i= n$ as
\be
\delta(N- \sum_i n_i) = \frac 1{2\pi} \int d\theta e^{i\theta(N-\sum_i n_i)},
\ee
and summing the geometric series after using the 
explicit expression (\ref{eigenfunctions}) for the $\omega_n$.

Instead of working  with the canonical ensemble we eliminate  
the constraint on the sum over the $n_i$ by working with the 
grand canonical ensemble. The grand canonical partition function
is defined by
\be
Z = \sum_n z^n Z_n= \prod_k (1+e^{-\beta \omega_k+\beta \mu}),
\ee
where $z= e^{\beta \mu}$ is the fugacity. 
The one particle density in the grand
canonical ensemble is given by
\be
\langle \rho(x) \rangle &=& \frac 1Z \sum_n z^n Z_n \rho(x)\nonumber \\
&=&\frac 1Z \sum_{n_i \in \{0,1\}} \sum_i n_i |\phi_i(x)|^2 e^{-\beta \sum_k
n_k(\omega_k-\mu)},
\ee 
which can be evaluated to be
\be
\langle \rho(x) \rangle = \sum_i |\phi_i(x)|^2 
\frac 1{1+ e^{\beta (\omega_i-\mu)}}.
\ee
The fugacity $z =e^{\beta \mu} $ is determined by the condition that
the total number of particles is equal to $n$, i.e.
\be
\sum_i  \frac 1{1+ e^{\beta (\omega_i-\mu)}} = n.
\label{chemfix}
\ee

The connected two-particle correlation function is defined by
\be
R_2(x,y) = \sum_{k\ne l}\langle \delta(x-x_k) \delta(y-y_l) \rangle
-\langle \rho(x) \rangle \langle \rho(y) \rangle 
\label{R2def}
\ee
It can be obtained from the $n$-particle matrix element (\ref{matrixelement})
 by integration
over all coordinates (eigenvalues) except $x_1$ and $x_2$. The remaining 
sum over 
$\omega_{\pi(1)}$ and $\omega_{\pi(2)}$ can be rewritten as
\be
\sum_{\omega_1< \cdots <\omega_N}\sum_{i\ne j}
\phi_{\omega_i}(x_1)\phi_{\omega_j}(x_2)\left [\phi_{\omega_i}(x_1)
\phi_{\omega_j}(x_2)
-\phi_{\omega_j}(x_1)\phi_{\omega_i}(x_2)\right ]
e^{-\beta \sum_k \omega_k}.
\ee
The diagonal term, $i=j$, is zero allowing us to include it in the summation.
The 
first term  
can then be identified as the square of the average particle 
density. It is the disconnected contribution to the two-particle
correlation function. The connected part of the two-particle
distribution function can thus be written as
\be
R_2(x_1, x_2) = -\frac 1{Z_n}\sum_{\omega_1< \cdots <\omega_n}
\sum_{i,j}
\phi_{\omega_i}(x_1)\phi_{\omega_j}(x_2)
\phi_{\omega_j}(x_1)\phi_{\omega_i}(x_2)
e^{-\beta \sum_i \omega_i}.
\ee
This correlation function does not include the contributions from 
the self-correlations of the eigenvalues.
 After all, our starting point was the joint distribution
of {\it different} eigenvalues.
In an occupation number representation $R_2(x_1, x_2) $ simplifies to
\be
R_2(x_1, x_2) = -\frac 1{Z_n}
\sum_{ n_1 + n_2 + \cdots =n} \left [\sum_{i} 
n_i \phi_{\omega_i}(x_1) \phi_{\omega_i}(x_2) \right ]^2  e^{-\beta\sum_k 
n_k\omega_l}.
\ee
Using this representation,
the two-particle density in the grand canonical ensemble is found to be
\be
R_2(x,y) = -\left |\sum_{i} \phi_i(x)\phi_i(y)
 \frac 1{1+ e^{\beta (\omega_i-\mu)}}\right |^2.
\label{r2grand}
\ee

Using similar manipulations as for  the canonical 
partition function and the corresponding one-particle density it is
possible to simplify  the exact analytical expression
for the  connected two-particle correlation function in the canonical ensemble,
\be
R_2(x,y) = - \sum_{k,l} |\phi_k(x)\phi_l(y)|^2 \sum_{r,s=0}^\infty
(-1)^re^{-4\beta\omega(kr+ls
-n(r+s) +\frac 12 (r+s)(r+s+1))}\prod_{p=N-r-s}^N(1-q^p).
\nonumber \\
\ee
This result can be used to compare the two ensembles but we will not address
this question in this article.

\section{Semiclassical Calculation}

In this section we calculate the microscopic spectral density and the 
two-point correlation function using semi-classical methods starting
from the expressions for the grand canonical ensemble derived in 
previous section. We are thus
interested in the region around $x=0$. Because of the hard edge at $x=0$
we cannot simply do a WKB approximation by replacing the wave functions
by plane waves but instead have to use Bessel functions. This follows
immediately from the wave equation for $x\to 0$, 
\be
\left [ -\partial_x^2 + \frac{4\nu^2 -1}{4x^2} \right ] \sqrt x J_\nu(kx) 
= k^2 \sqrt x J_\nu(kx).
\ee 
Alternatively, one can exploit the asymptotic relation between 
Laguerre polynomials
and Bessel functions.

\subsection{The Microscopic Spectral Density}

Taking into account the normalization of the Bessel functions,
$\int_0^\infty kdk\sqrt{xx'} J_\nu(kx)J_\nu(kx') = \delta(x-x')$, 
to fix the constants in the integration measure, 
we arrive at the following expression for the single particle density
\be
\langle \rho(x) \rangle = 
\int_0^\infty kx dk \frac{J_\nu^2(kx)}{1+e^{\beta k^2 -\beta\mu + 
\beta\omega^2 x^2}}.
\label{semirho}
\ee
The chemical potential is determined by the condition 
$\int dx \rho(x) = n$.
Since this integral is over all $x$, the
use of the semi-classical expressions
for the wave functions is not justified
but instead we have to rely on the exact wave functions.
In the limit, $\beta\mu \gg \omega$, 
the sum over $i$ in (\ref{chemfix}) 
can be replaced be an integral which can be performed 
 analytically resulting in
\be
e^{(2\nu+2)\beta\omega - \beta \mu} = \frac 1{e^{4n\beta \omega} - 1}.
\ee
In the limit $n\beta\omega \gg 1$ the semi-classical expression for the
spectral density is thus given by
\be
\langle \rho(x) \rangle= \int_0^\infty kx dk \frac{J_\nu^2(kx)}
{1+e^{\beta k^2 +\beta\omega^2 x^2-(4n-2\nu-2)\beta\omega}}.
\label{nxsc}
\ee
In this limit the semiclassical expression (\ref{semirho}) 
leads to the correct value
of the chemical potential.
Below we will show for $n\beta\omega \gg 1$ and finite $x$ (in units
of the average level spacing)
 the term $ \beta \omega^2 x^2 $ can be neglected relative to $4n\beta\omega$. 

An estimate for the average spectral density near zero but many level
{\it spacings} away from $x=0$ is obtained by using  
 the leading order asymptotic expansion of the
Bessel functions and calculating the integral in (\ref{nxsc}) 
in the limit of a degenerate Fermi-gas.
This results in
\be
\bar \rho= \frac {\bar k} \pi,
\ee
where  $\bar k \approx \sqrt{4n\omega}$ is the ``radius''of the Fermi-sphere.
 At fixed $h$ in the limit $n\gg 1$
we have
\be
\omega \approx \Sigma^2 n \qquad {\rm and} \qquad
\beta \omega \approx \frac 1{2hn}.
\ee
Using these results and invoking the Banks-Casher formula
the parameter $\Sigma$ can be identified as
the chiral condensate,
\be
 \lim_{n\rightarrow \infty} \frac {\pi \bar \rho}{2n}= \Sigma.
\ee

We will now show that our approximations are self-consistent.
The condition that we are close to the degenerate Fermi gas can be written
as $\delta k /\bar k \ll 1$. We thus have to impose the requirement that
\be
\frac{\delta k}{\bar k} \sim \frac 1{2\beta {\bar k}^2} =\frac h4 \ll 1.
\ee
The conditions $kx\gg 1$ and $\beta\omega^2 x^2\ll 4n\beta \omega$ can be
combined into
\be
\frac 1{4n} \ll \omega x^2 \ll 4n,
\ee
 or, in units
of the average level spacing, $x = u/\bar \rho $, the range of
validity of the above asymptotic results is given by
\be
\frac 1{4n} \ll \frac {u^2\pi^2}{4n} \ll 4n.
\label{range}
\ee
Because of the second inequality it is justified to  neglect the term
$\beta \omega^2 x^2$ which we will do in the remainder of this section.

By partial integration the expression  (\ref{nxsc}) for the spectral
density can be rewritten as
\be
\langle \rho(x)\rangle = 
\int_0^\infty dk \frac{\beta xk^3\left [ J_\nu^2(kx)- J_{\nu+1}(kx)
J_{\nu-1}(kx)\right ]}{4\cosh^2 \frac \beta 2(k^2 -4n\omega)}.
\label{onep}
\ee
Using the Banks-Casher formula one finds that the chiral condensate
depends on $h$. The leading order correction is given by
\be
\frac{\Sigma(h)}{\Sigma} = 1- \frac{{\pi}^2 }{96} h^2 + \cdots.
\ee
The corresponding spectral density will be denoted by 
$\bar \rho(h)\equiv 2n\Sigma(h)/\pi$.
The microscopic spectral density is then given by
\be
\rho_s(u) &=& \lim_{n\rightarrow \infty}
\frac 1{2n\Sigma(h) } \rho(\frac u{2n\Sigma(h)}),\nonumber \\
&=& \frac{\Sigma}{\Sigma(h)}\int_0^\infty dk
\frac{ k^2 \rho_s^0(ku \Sigma/\Sigma(h))}{ 
h\cosh^2\frac {(k^2 - 1)}h}.
\label{microex}
\ee
where $\rho_s^0$ is the microscopic spectral density for the chGUE 
\cite{VZ,V}
\be
\rho_s^0(u) = \frac u2 \left[ J_\nu^2(u)- J_{\nu+1}(u)
J_{\nu-1}(u)\right ].
\label{micro2}
\ee
The interpretation is clear. The oscillations in the microscopic spectral
density due to the Bessel functions 
are smeared out over a distance $hu$ by the integration over $k$.
The oscillations are thus visible up to a distance of $u \sim 1/h$.

\vspace*{2cm}
\begin{figure}[ht]  
\begin{center}
\epsfig{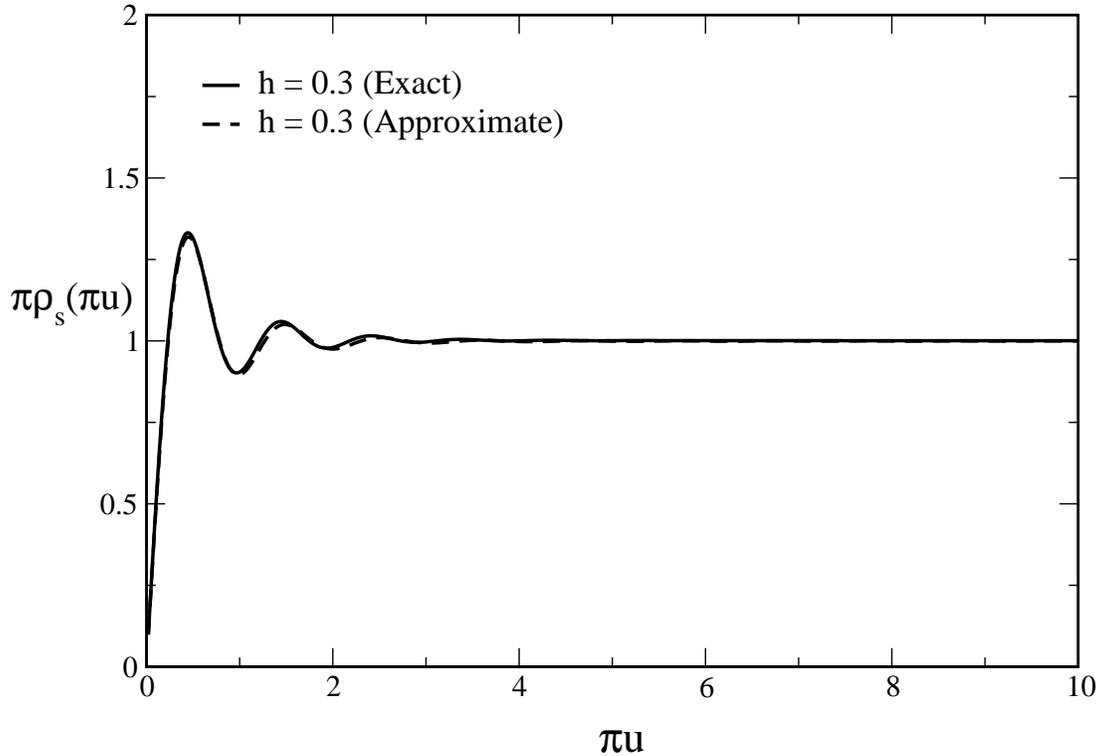}
\caption{The microscopic spectral density $\pi\rho_s(\pi u)$ versus $\pi u$ for
a value of $h =0.3$. The exact result (full curve) and the approximate
result (dashed curve) nearly coincide. }
\label{callisto1}
\end{center}
\end{figure}

In the limit of small $h$ the main contribution to the integral comes from
the region around the Fermi-surface. In this limit we can derive an approximate
formula correct to order $h^2$ at fixed $uh$. 
To this end
we change integration
variables in (\ref{microex}) according to $k = 1+ht$ neglecting terms 
that are sub-leading
in $h$. This results in
\be
\rho_s(u) 
= \int_{-\infty}^\infty dt \frac{\rho_s^0(u+hut)}{\cosh^2(2t)}.
\label{integral}
\ee
Next we Taylor expand the microscopic spectral density
 as follows
\be
\rho_s^0(u+hut) &=& \rho_s^0(u) \nonumber\\
&+& \frac 12 hut\left [ J_{\nu}^2 +J_{\nu+1}J_{\nu-1})
\right ]\nonumber\\
&+&\frac 12 \frac {(hut)^2}{2!}  \left [4 J_{\nu} J_{\nu-1}+
\frac{f_1}{u}\right ]\nonumber \\
&+& \frac 12\frac {(hut)^3}{3!}\left[ 4(J_{\nu-1}^2- J_{\nu+1}^2) 
 +\frac {g_1}{u}+\frac {g_2}{u^2})\right ]\nonumber \\
&+& \frac 12\frac {(hut)^4}{4!} \left [-4^2 {J_{\nu}J_{\nu-1}}
 +\frac {h_1}{u^1}+\frac {h_2}{u^2}+\frac {h_3}{u^3})\right ],
\ee
where the terms that will be neglected are denoted by $f_i$, $g_i$,
$h_i$, etc.. At small values of $u$ these terms are of order $h^2u$ whereas
for large $u$ they are  suppressed by order $1/u^2$ 
(notice the factor $u$ in (\ref{micro2})).
By inspection one easily finds that the neglected terms are at most of
order $h^2$ independent of the value of $u$. The leading order terms can
be easily resummed to
\be
\rho_s^0(u+hut) = \rho_s^0(u) -  \frac 12 J_{\nu}(u)J_{\nu-1}(u) 
[\cos (2hut) -1 ] +O(h^2) +\,\,\, {\rm terms \,\,odd\,\,in\,\,}t.
\ee
The integral over $t$ in (\ref{integral}) can be performed analytically
resulting in
\be
\rho_s(u) = \rho_s^0(u) -  \frac 12
J_{\nu}(u)J_{\nu-1}(u)\left [ \frac {\pi hu}{
2 \sinh(\pi hu/2)} -1 \right ] +O(h^2).
\ee
In Fig. 1 we compare the exact expression for the microscopic spectral density
(\ref{microex}) to this approximate formula. We observe that even for
 a value of $h$ as large as $h = 0.3$ the two results are very close.

\subsection{Two-Point Function}

 In this subsection we derive a semi-classical expression for 
the two-point correlation function. To this end
the wave functions in the expression (\ref{r2grand}) for the
connected two-point correlation function
are replaced by Bessel functions,
\be
R_2(x,y) = -\left |\int_0^\infty k dk \frac{ \sqrt {xy}J_\nu(kx)J_\nu(ky)}
{1+e^{\beta k^2 -4n\beta\omega}}\right |^2.
\label{twop}
\ee
By partial integration with respect to $k$ the correlation function can be
expressed as
\be
R_2(x,y) = -\left |\frac{\bar \rho}h 
\int_0^\infty k^2 dk \frac{K(x\pi\bar \rho k, y\pi \bar \rho k)}{\cosh^2((k^2-1)/h)}
\right |^2,
\ee
where $K(x,y)$ is the two-point kernel for the chGUE given by \cite{VZ,V}
\be
K(x,y) = \sqrt{xy} \frac{xJ_{\nu+1}(x)J_{\nu}(y)-yJ_{\nu}(x)J_{\nu+1}(y)} 
{x^2-y^2}.
\ee
To  order  $h^2$ the correlation function can be simplified to
\be
R_2(x,y) = -\left |{\bar \rho} 
\int_{-\infty}^\infty  dt \frac{K(x\pi\bar 
\rho (1+ht), y\pi \bar \rho (1+ht)}{\cosh^2 2t}
\right |^2 + O(h^2),
\label{r2t}
\ee
In the same way as for the one-point function we now will
derive an approximate formula for
the two-point function correct to order $h^2$ at fixed value of $uh$. This can
be done conveniently by using the following summation formula  
\be
 J_{\nu+1}(x+xh) J_\nu(y+yh)&\pm& J_{\nu}(x+xh) J_{\nu+1}(y+yh)\nonumber \\
      &\approx&
[ J_{\nu+1}(x) J_\nu(y)\pm J_{\nu}(x) J_{\nu+1}(y)] \cos[(x+y)h]\nonumber \\
&+&[J_{\nu}(x)J_{\nu}(y)\mp J_{\nu+1}(x)J_{\nu+1}(y)] \sin[(x+y)h].
\ee
This formula has been derived by means of a Taylor expansion and a subsequent
resummation employing the following approximate derivative formulas 
\be
\partial^{2n}_k 
[J_{\nu+1}(kx) J_\nu(ky)&\pm& J_{\nu}(kx) J_{\nu+1}(ky)] \nonumber
\\ &\approx& 
(-1)^n (x\pm y)^{2n} [J_{\nu+1}(kx)J_{\nu}(ky) \pm J_{\nu}(kx)J_{\nu+1}(ky)], 
\ee
and 
\be
\partial^{2n-1}_k 
[J_{\nu+1}(kx) J_\nu(ky)&\pm& J_{\nu+1}(kx) J_\nu(ky)] 
\nonumber \\ &\approx&
(-1)^{n+1}(x\pm y)^{2n-1} [J_{\nu}(x)J_{\nu}(y)\mp J_{\nu+1}(x)J_{\nu+1}(y)].
\ee
They have been obtained by means of recursion 
relations for Bessel functions neglecting terms that are suppressed by 
order $1/x$ or $1/y$. One can easily show that the combined powers
of $x$ and $y$ in the prefactor is always larger than the combined powers of
$x$ and $y$ that have been neglected. 
Since only even terms in $t$ contribute to the integral in 
(\ref{r2t}) our final
result for the two-point function, correct to order $h^2$, is given by 
\be
R_2(x,y) =  && -\bar \rho^2 \left | \frac{\pi^2 h \bar \rho  \sqrt{ xy}}8 
\left [\frac{ J_{\nu+1}(x\pi \bar \rho) J_\nu(y\pi \bar \rho)
+ J_{\nu}(x\pi \bar \rho) J_{\nu+1}(y\pi \bar \rho)}
{\sinh((x+y)\pi^2 h \bar \rho/4) }  \right ] \right . \nonumber \\
&+& \hspace*{1cm}\left .  \frac{\pi^2 h \bar \rho \sqrt{ xy}}8 
\left [\frac{ J_{\nu+1}(x\pi \bar \rho) J_\nu(y\pi \bar \rho)
- J_{\nu}(x\pi \bar \rho) J_{\nu+1}(y\pi \bar \rho)}
{\sinh((x-y)\pi^2 h\bar \rho/4) }  \right ]
 \right |^2
\label{r2an}
\ee
In the limit $x, \, y \gg |x-y|$ the analytical result for the 
two-point function of the model of Moshe, Neuberger and Shapiro \cite{Moshe} 
is recovered from
the leading order asymptotic expansion of the Bessel functions.
For unitary invariant ensembles, it can be shown that  the result 
of \cite{Moshe} for
critical statistics and Wigner-Dyson statistics are the only two
possibilities \cite{kanpri}. At this moment it is not clear whether this
argument can be extended to the chiral unitary ensembles as well. 

\vspace*{2cm}
\begin{figure}[ht]  
\begin{center}
\epsfig{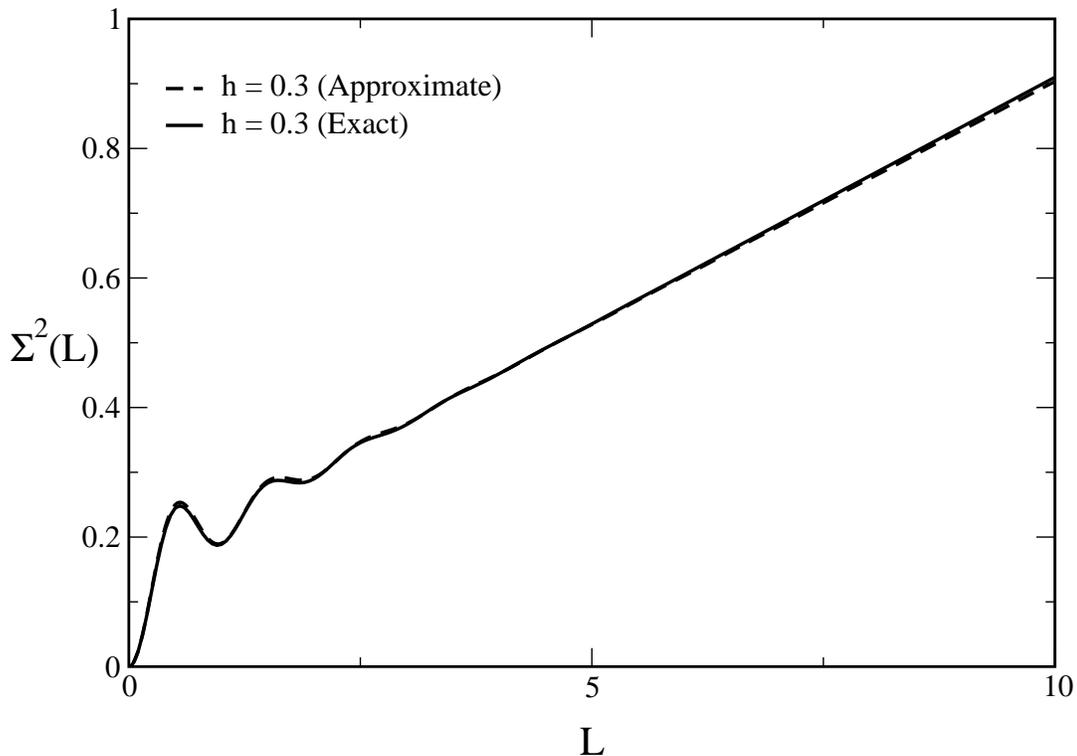}
\caption{The number variance $\Sigma^2(L)$ versus $L$ for
a value of $h =0.3$. The exact result (full curve) and the approximate
result (dashed curve) nearly coincide.} 
\label{callisto2}
\end{center}
\end{figure}

The number variance of the eigenvalues near $x=0$ is obtained by
integrating the two-point correlation function including 
the self-correlations
\be
\Sigma^2(L) = \int_0^{L /\bar \rho(h)} dx\int_0^{L /\bar \rho(h)} dy\left [
\delta(x-y) \langle \rho(x) \rangle + R_2(x,y) \right ].
\ee
We study its asymptotic behavior  in the limit $L \rightarrow \infty$. 
Starting from 
the expressions (\ref{onep}) and (\ref{twop}), $\Sigma^2(L)$ 
can be simplified by
means of an orthogonality relation 
for Bessel functions. To leading order in $h$ we find
\be
\Sigma^2(L) &=& \int_0^{L /\bar \rho(h)} xdx
\int_0^\infty k  dk \frac{J_\nu^2(kx)}
{4\cosh^2\frac{\beta }2(k^2 -4n\omega)}\nonumber \\
&=& \frac {L}{\bar \rho 2\pi \beta \bar k} = \frac h{4} L.
\ee
The same asymptotic  result can be derived from the analytical result
(\ref{r2an})  (In this case the average spectral density
 does not depend on $h$ (see \ref{integral}).).  
 Such linear term, first proposed in {\cite{Altshuler}},  is believed to
be characteristic for  universal critical statistics \cite{shklov} valid at 
the mobility edge and has been related to the
multifractality of the wave functions \cite{huck,chalker,chalker2,krats}.

In Fig. 2 we compare the number variance derived from the 
approximate analytical 
result (\ref{r2an}) and from the exact result (\ref{twop}).
Clearly, even for a value of $h$ as large as 0.3 the two curve are 
barely distinguishable.

The  asymptotic linear behavior of the number variance seems to be contradicted
by the sum rule
\be
\int_{-\infty}^\infty  dy[\delta(x-y) \langle \rho(x) \rangle + R_2(x,y) ] = 0.
\ee
The resolution
of this paradox \cite{kra1,kra2} is probably best illustrated by considering
the Poisson ensemble for $n$ uncorrelated eigenvalues with average 
spacing $\bar \rho$.
To satisfy the sum-rule we have that $R_2(x,y) = \bar \rho/n$ 
instead of zero for uncorrelated eigenvalues
resulting in the number variance $\Sigma_2(L) = L -L^2/n$. We  conclude that
an asymptotic linear behavior is possible if the thermodynamic limit is taken
before the limit $L\rightarrow \infty$.

To make contact with the partially quenched effective partition function, 
for which
the number of subsequent eigenvalues 
around zero that are correlated according to the chGUE 
scales as $\sqrt n$ \cite{Vplb,james,OTV,DOTV}, we have to scale $h$ as
$h \to h/\sqrt n$. In this 
limit microscopic universality 
\cite{SV,brezin,Damgaard,Sener1,GWu,Seneru} is recovered
for the interpolating chiral unitary ensemble.  

\vspace*{1cm}
\begin{figure}[ht!]  
\begin{center}
\epsfig{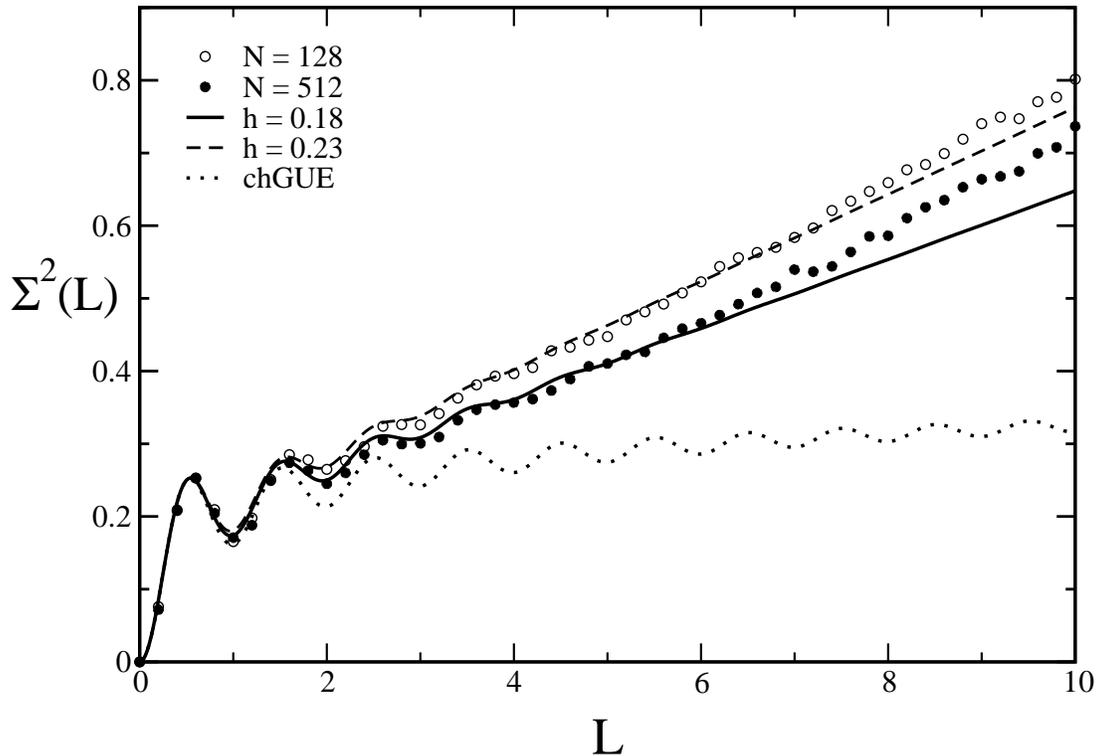}
\caption{The number variance $\Sigma^2(L)$ versus $L$
 for Dirac eigenvalues 
of instanton gauge field configurations (points) and our interpolating
random matrix model. The total number of instantons and the parameter $h$
of our model are given in the label of the figure.}
\label{callisto3}
\end{center}
\end{figure} 

\section{Comparison with Instanton Simulations}
Spectral correlations have been studied in great detail for both lattice
QCD simulations \cite{Vplb,berbenni,many,Edwa99b,Damg99a,Berg99,Farc99a}
and instanton-liquid simulations \cite{james}
(see \cite{tilojac} for a recent review). 
In lattice QCD
they were studied by means of the disconnected scalar susceptibility 
\cite{Berb99}, and complete
agreement with partially quenched chiral perturbation theory was found.
In particular, it  was shown that the number of subsequent 
eigenvalues around zero
described by chRMT scales as $F^2 \sqrt V$. In instanton simulations 
a weaker volume dependence of the number of such eigenvalues 
was observed 
suggesting an approach to a critical point similar to a
localization transition. Indeed the 
 multifractality index of the fermionic wave functions was found to be nonzero.
We thus compare the instanton data
 with the model  in previous section at fixed value of the parameter $h$.  
Results for the number variance, $\Sigma^2(L)$ versus $L$ are shown in Fig. 3. 
The closed and open circles represent results \cite{james}
for the eigenvalues of the Dirac operator with field configurations
given by an ensemble of instantons and an equal number
of anti-instantons with a total density of 1 $fm^4$. 
The total number of (anti-)instantons is given in the label 
of the figure. In the same figure we show the result for the chGUE 
(dotted curve)
and results for the model (\ref{twop}) for $h = 0.18$ (full curve)
and $h = 0.23$  (dashed curve). We observe that both the slope
and the range of agreement with 
the chGUE curve only shows  a week volume dependence. Outside this domain the
data show a linear $L$-dependence. 
Both features are nicely reproduced by the critical Random Matrix Model.
The slightly positive curvature of the instanton data might be
a remnant of the  $L^2$ dependence predicted for the 
Altshuler-Shklovsky domain \cite{Altshuler}.  

\section{Conclusions}
   
We have analyzed a chiral random matrix model that 
interpolates between the chGUE and the
chiral Poisson ensemble. This model is a generalization 
of a model originally proposed
by Moshe, Neuberger and Shapiro \cite{Moshe}.
It has been  mapped onto a gas of non-interacting fermions
and was solved by means of statistical mechanics methods. To leading order in
the deviation from the chGUE we have obtained compact analytical expressions for
the microscopic spectral density and the two-point level correlation function.
We have shown that this critical chiral random matrix
 model provides a good description of the level correlations
of Dirac eigenvalues for gauge field configurations given by a liquid of instantons.

The number variance  of the critical chiral random matrix
model shows a linear
$L$-dependence for a large  $L$ (in units of the average level spacing)
whereas it coincides with the chGUE result for small values of $L$. The 
characteristic feature is that the transition point between these two
domains is stable in the thermodynamic limit. 
This situation is very different for a non-linear $\sigma$-model description
of disordered systems where this transition point or 
the Thouless energy is determined by the competition between the mass term
and the kinetic term. In that case  one finds the scaling behavior 
$E_c \sim D/L^2_s$ with $D$ the diffusion constant
and $L_s$ the linear size of the sample.  
The theoretical reason for a scale independent dimensionless 
conductance (i.e. the Thouless energy in units of the average level spacing)
is that the localization length diverges 
at a critical value of the disorder.
In the approach to this limit the diffusion 
constant has to become scale dependent. 
 If $E_c$, in units of the average level spacing, becomes scale independent 
the diffusion constant has to be scale dependent leading to 
a multi-fractal scaling of the wave functions. 

The weak volume dependence and the linear number variance
observed in correlations of eigenvalues of the QCD Dirac operator
with instanton liquid gauge field configurations   
suggests that we are dealing with a critical system close to 
a localization transition. Indeed the same numerical simulations suggest
a small nonzero multifractality index of the wave-functions.
On the other hand, the dimensionless conductance found in lattice QCD 
simulations scales according
to our expectations from chiral perturbation theory. At this moment 
we do not have a good explanation for this discrepancy. It could simply
be that the expected scaling behavior is only recovered for 
much large volumes in instanton simulations.
Indeed, a very
slow approach to the thermodynamic limit 
has been found for other quantities such a 
quenched chiral logarithms.
Clearly, more work has to be done to resolve this issue.

\vskip 0.5 cm
{\bf Acknowledgements}
\vskip 0.5cm

This work was partially supported by the US DOE grant
DE-FG-88ER40388. One of us (A.M.G.) was supported by ``laCaixa Fellowship
Program''.  J.J.M.V. thanks the Institute for Nuclear Theory at
the University of Washington for its hospitality and partial support
during the completion of this work. Dominique Toublan is thanked
for a critical reading of the manuscript.

\end{document}